\documentclass[usenatbib,usegraphicx]{mn2e}
\usepackage{amsmath}
\usepackage{amssymb}
\usepackage{mathrsfs}
\usepackage{color}
\usepackage{subfigure}

\newcommand{\rmd}{{\mathrm{d}}}
\newcommand{\Msun}{\:{\rm M}_{\odot}}

\newcommand{\rhalf}{{\rm r}_{_{1/2}}}

\newcommand{\rcore}{{\rm r}_{_{\rm core}}}
\newcommand{\rlimit}{{\rm r}_{_{\rm lim}}}
\newcommand{\pcore}{\rho_{_{\rm core}}}
\newcommand{\rcut}{{\rm r}_{_{\rm cut}}}

\newcommand{\Mhalf}{{\rm M}_{_{1/2}}}

\newcommand{\LCDM}{\Lambda{\rm CDM}}
\newcommand{\beq}{\begin{equation}}
\newcommand{\eeq}{\end{equation}}

\newcommand{\Vmax}{V_{\rm max}}
\newcommand{\rmax}{{\rm r}_{_{\rm max}}}

\begin{document}

\title[Concentrations $\&$ Cores in Milky Way dSphs]{Dark matter
  concentrations and a search for cores in Milky Way dwarf satellites}

\author[Wolf \& Bullock]
{Joe Wolf \& James S. Bullock \\
Center for Cosmology, Department of Physics and Astronomy, University of California, Irvine, CA 92697\\
wolfj@uci.edu, bullock@uci.edu\\
}

\date{\today}

\maketitle

\begin{abstract}

We investigate the mass distributions within eight classical Milky Way dwarf
spheroidal galaxies (MW dSphs) using an equilibrium Jeans analysis and we compare our results
to the mass distributions predicted for subhalos in dissipationless $\LCDM$ simulations.
In order to match the dark matter density concentrations predicted, the stars in these galaxies must
have a fairly significant tangential velocity dispersion
anisotropy ($\beta \simeq -1.5$).
For the limiting case of
an isotropic velocity dispersion ($\beta = 0$), the classical MW dSphs predominantly prefer to live in halos that are
less concentrated than $\LCDM$ predictions. 
We also investigate whether the dSphs prefer to live in halos with constant
density cores in the limit of isotropic velocity dispersion. Interestingly, even in this
limit,  not all of the dSphs
prefer large constant-density cores: the Sculptor dSph prefers a cusp
while Carina, Draco and Leo I prefer cores. The other four dSphs do
not show a statistically significant preference for either cuspy or cored profiles.
Finally, we re-examine the
hypothesis that the density profiles of these eight MW dSphs can be quantified by a common dark matter halo.
\end{abstract}


\section{Introduction}
\label{sec:intro}

The currently favored cold dark matter cosmology ($\LCDM$) has had
much success in reproducing the large scale structure of the universe
\citep[see][and references therein]{WMAP7}. However, on smaller scales
there are some possible discrepancies
\citep[e.g.][]{FloresPrimack94,Moore94,Klypin99, Moore99,BK11}. One robust
prediction from {\em dissipationless} $\LCDM$ simulations is that the
inner density profile of dark matter halos should be fairly cuspy at
small radii\footnote{Interestingly,  \citet{Penarrubia10} 
  showed that dSphs are more likely to survive interactions with the
  MW disk if they  are cusped rather than cored.}, with $\rho(r)
\propto r^{-\gamma}$ and  $\gamma \simeq 1$
\citep[][]{Dubinski91,nfw,VL2, Aquarius}. In contrast, some
observed rotation curves from dark matter dominated galaxies prefer
fits with $\gamma$ closer to 0 than 1 \citep[e.g.][]{deBlok01,
Simon03, Simon05, Kuzio08, deBlok10}. 
Even when fits to the rotation curves are forced to $\gamma = 1$ models,
the resulting profiles are often less dense and less centrally-concentrated
than predicted from dissipantionless $\LCDM$ simulations \citep[e.g.,][]{Alam02}.
 At face value, the central density problem and
cusp-core problem provides a challenge to cold dark matter theory.   

Milky Way dwarf spheroidal galaxies (MW dSphs)
provide particularly interesting dark matter laboratories.
They include the most dark matter dominated systems  
known and are close enough to enable star-by-star kinematic studies
\citep[e.g.][hereafter W10]{Walker07, SG07, Strigari08, Geha09, Lokas09, Kalirai10,
Martinez10, Simon11, Willman10, Willman11, Wolf10}. 
Unfortunately, because these objects are dispersion-supported their
unknown stellar velocity anisotropy makes it  
difficult to determine their mass distributions using a simple projected
Jeans analysis \citep[e.g.][W10]{Strigari07a}.   

In this paper, we ask whether a simple class of equilibrium models
prefer solutions with dark matter densities that are consistent with $\LCDM$ expectations,
and we also explore whether the solutions allow 
constant dark matter density in the inner
parts, in conflict with naive expectations of $\LCDM$ theory.

Our work is motivated in part by the earlier study by \citet{Gilmore07}
(hereafter G07) where the results of an isotropic spherical Jeans
analysis of the dispersion profiles of six of the eight brightest
MW dSphs -- Carina, Draco, Leo I,
Leo II, Sextans, and Ursa Minor -- were presented. The main result of
G07 was that solutions to the Jeans equation for these systems prefer
dark matter models with an ``approximately constant density out to
some break radius".  In this paper, we include all eight bright MW
dSphs (excluding the tidally-disrupting Sagittarius dSph). Our results
are different and do not lend support to the claim that all MW dSphs
inhabit halos with large constant density cores. 
We look at the full probability distribution of the inner slope of the
dark matter density profiles and find the spread in allowed values to
be large enough that we cannot make a definitive statement for half of
these dSphs. Among the other half, our analysis shows that Sculptor
strongly prefers a cuspy halo. 

We pay careful attention to the degeneracy of the dark matter density
profile slope and the inner density profile of stars in our
analysis. The inner density profile of the tracer stars (i.e. those with velocity
measurements), as well as the stellar density profile derived from
photometry, are uncertain in the inner parts. This  uncertainty
introduces significant uncertainty in the determination of the dark
matter density profile -- a point highlighted recently in
\citet{Strigari10} (hereafter S10). We note that this degeneracy is
significant, even in comparison to the well-known degeneracy of the mass profile
with the unknown velocity dispersion anisotropy.   

Our conclusions are conservative in the sense that we force the inner
light profiles to follow King (1962) models, which tend to bend sharply
to cored light distributions as $r \rightarrow 0$ and thus favor
more core-like mass densities (see \S 2). King profiles are not unique
fits for these systems, and by allowing a broader range of stellar
profile shapes a wider range of density profile slopes become
possible (S10). 

If there are constant density cores at the centers of small dark
matter halos it could signal the existence of non-standard dark  
matter particles \citep[e.g.][]{HD00, Avila01, Kaplinghat05, 
Strigari07c}. A modified dark matter solution would be particularly  
attractive if these cores were ubiquitous and of approximately the
same density. Density cores might alternatively arise
from baryonic physics that alters the dark matter distributions
compared to dissipationless simulations  \citep[as demonstrated
explicitly by][]{Mashchenko08, Governato10, Pasetto10, Oh11}.    The
feedback processes that give rise to such cores rely on potentially
stochastic details like the fraction of baryons that have been
converted into stars \citep[e.g.][]{Brook11}. Therefore, if the cores
are created by feedback processes, one might expect that the central
density structure of dark matter halos will demonstrate significant  
variations from halo to halo, as seems to be the case for low surface
brightness galaxies \citep{Kuzio10}. 

Lastly, we further explore the anisotropy-slope degeneracy in order
to address the question of what anisotropies are needed in order for
simulations to reproduce these MW dSphs, if they are to live in
$\LCDM$-like halos.

In the next section we provide a discussion of the
degeneracies inherent in the use of Jeans modeling to constrain 
density profile slopes, 
even under the assumption of isotropy. We present our analysis
and results in \S 3, where we discuss the mass modeling for 
each dSph, the question of a common halo for all the dSphs,
the most probable stellar velocity dispersion anisotropies 
required to fit NFW profiles to the dSphs and a comparison of 
concentrations to those seen in recent $\LCDM$ simulations.
We also discuss both the requirements for our cuspy density profile
fits to be physical when isotropy is assumed and the complexities 
in the stellar population that we do not take into account in this work. 
We summarize our results in \S 4.  

\section{Expectations from the Jeans equation}
\label{sec:expect}
\begin{table*}
{\bf Table 1:} Derived parameters of MW dSphs when modeled under the assumption of isotropy.\\
\vskip 0.1 cm
\begin{tabular}{lcccccccccc}
\label{tab:main}
Galaxy & $\rhalf$ [pc] & $C_{\rm King}$ & $\kappa_\star (\rhalf)$ & \vline & $\gamma$($r \rightarrow 0$ pc) & $\gamma$(100 pc) & $\gamma(\rhalf)$ & \vline & log$_{10}(\pcore \, \rcore \, \Msun^{-1} {\rm pc}^2$) & $\rcore$ [pc] \\ \hline \hline
Carina         &  334 & 3.3 & 1.9 & \vline & $0.0^{+0.5}$            & $0.4^{+0.3}_{-0.2}$ & $0.4^{+0.4}_{-0.3}$& \vline& $1.2^{+1.2}_{-0.1}$ & -- \\
Draco          &  291 & 5.9 & 1.2 & \vline & $0.0^{+0.4}$            & $0.3^{+0.3}_{-0.2}$ & $0.5^{+0.3}_{-0.1}$& \vline& $2.0^{+0.4}_{-0.3}$ & -- \\
Fornax        &  944 & 5.2 & 0.8 & \vline & $0.7^{+0.3}_{-0.4}$  & $0.8^{+0.2}_{-0.4}$ & $1.4^{+0.3}_{-0.2}$& \vline& $1.1^{+0.04}_{-0.03}$ & $790 ^{+130}_{-110}$ \\
Leo I           &  388 & 1.9 & 2.7 & \vline & $0.0^{+0.4}$            & $0.2^{+0.3}_{-0.1}$ & $0.4^{+0.4}_{-0.2}$& \vline& $2.3^{+0.5}_{-0.6}$ & -- \\
Leo II          &  233 & 3.5 & 2.2 & \vline & $0.5^{+0.6}_{-0.3}$ & $0.8^{+0.5}_{-0.4}$ & $1.2^{+1.6}_{-0.5}$& \vline& $1.4^{+1.7}_{-0.1}$ & -- \\
Sculptor      &  375 & 13 & 0.5 & \vline & $1.1^{+0.1}_{-0.5}$ & $1.2^{+0.1}_{-0.4}$ & $1.3^{+0.2}_{-0.1}$& \vline& $0.61^{+0.27}_{-0.04}$ & $410 ^{+80}_{-70}$ \\
Sextans      & 1019 & 9.6 & 2.4 & \vline & $0.7^{+0.4}_{-0.4}$ & $0.8^{+0.4}_{-0.3}$ & $1.5^{+0.5}_{-0.3}$& \vline& $0.61^{+0.27}_{-0.15}$ & $660 ^{+280}_{-200}$ \\
Ursa Minor & 588   & 4.4 & 1.1 & \vline & $0.3^{+0.6}_{-0.2}$ & $0.5^{+0.4}_{-0.3}$ & $1.0^{+0.5}_{-0.4}$& \vline& $1.5^{+0.7}_{-0.1}$ & -- \\
\end{tabular}\\

\vskip 0.2 cm
Note: Mean values of deprojected half-light radii ($\rhalf$) are from
Table 1 of W10 while the mean King concentrations 
$C_{\rm King} \equiv (\rlimit / {\rm r_{_{King}}})$ and $\kappa_\star$ values 
are derived from the photometric properties listed in Table 1 of W10.  Here $\kappa_\star (\rhalf)$ 
quantifies the curvature of the light profile at $\rhalf$, as defined by Equation \ref{eq:kappa}.  As discussed in the text, 
galaxies whose light profiles have smaller values of the curvature will tend to
favor cuspier overall density profiles. 
The total density log-slopes  $\gamma(r) =  - \rmd \ln \rho / \rmd
\ln r$ in the next three columns are derived when
utilizing Equation \ref{eq:rhor}. The density parameters in the last
two columns are derived by utilizing the Burkert profile (Equation \ref{eq:Burkert}). 
Errors correspond to 68\% likelihood values. 
\end{table*}

The spherical Jeans Equation relates the integrated mass of a spherically symmetric, dispersion-supported, collisionless system 
to its tracer velocity dispersion and tracer number density $n_{\star}(r)$, under the assumption of dynamical equilibrium:
\begin{equation}
\label{eq:mass-jeans}
M(r) = \frac{r \: \sigma_r^2}{G} \left (\gamma_\star+\gamma_\sigma - 2\beta \right ).
\end{equation}
Here $\sigma_r(r)$ is the radial velocity dispersion of the tracers
and $\beta(r) \equiv 1- \sigma_t^2 / \sigma_r^2$ quantifies the
tangential velocity dispersion.  The additional terms quantify radial
gradients in the stellar distribution 
 $\gamma_{\star} \equiv - \rmd \ln n_\star / \rmd \ln r$ and velocity dispersion profile $\gamma_{\sigma} \equiv - \rmd \ln \sigma_r^2 / \rmd \ln r$. 
While $\beta$, and to a lesser extent $\gamma_\sigma$, are difficult to determine empirically, $\gamma_\star$ is in principle constrained by
observations. If we assume a form of the anisotropy then $\gamma_\sigma$ can be solved for directly \citep[see][and Equations A7-9 in W10]{Mamon10},
and thus the Jeans equation immediately provides a path towards direct mass constraints. If we further assume that the stellar system
is isotropic ($\beta=0$), then our analysis simplifies considerably.

However, even with the assumption of isotropy, constraints on the slope of the mass density profile  $\gamma \equiv  - \rmd \ln \rho / \rmd \ln r$ will 
require knowledge of the second derivates of both $n_\star(r)$ and $\sigma_r(r)$, which will necessitate exquisite data sets. 
Typically, $n_\star(r)$ is assumed to follow an empirically-motivated functional form, without significant concern for the allowed freedom of 
its detailed curvature at small radius. For purposes of comparison, we will follow G07 and assume that $n_\star(r)$ tracks a \cite{King62} profile.
As we will now demonstrate, this is conservative in the sense that it forces a sharply flattening inner slope for the tracer stars, which in turn favors a 
flatter density slope $\gamma$ at small $r$ than might otherwise be allowed. Indeed, S10 found that most of the available 
photometric data for the classical dSphs can be fit with divergent 3D number densities that asymptote slowly to power-laws.  
In this sense, the inner density slopes derived in this paper are biased to favor more core-like behavior.

One can understand how the shape of $n_\star(r)$ affects core vs. cusp determinations by considering a proxy for the density profile slope 
 $\gamma \simeq \aleph \equiv 3 - \rmd \ln M / \rmd \ln r$.  For $\beta=0$,  Equation \ref{eq:mass-jeans} implies 
$\aleph = 2 + \gamma_\sigma - (\gamma_\star^\prime + \gamma_\sigma^\prime)/(\gamma_\star+\gamma_\sigma$), where the primes indicate 
derivatives with respect to $\ln r$.  If we define 
\beq
 \label{eq:kappa}
\kappa_\star \equiv \gamma_\star^\prime/\gamma_\star
\eeq
 in order to
quantify the curvature of the light profile, we can rewrite the density slope proxy as
 \beq
 \label{eq:approx}
\aleph = 2 + \gamma_\sigma - \gamma_\sigma^\prime \, \left(\gamma_\star + \gamma_\sigma \right)^{-1}\, - \kappa_\star \, 
\left(1 + \gamma_\sigma/\gamma_\star \right)^{-1}  \,   .
\eeq
At radii near the the deprojected half-light radius $r \simeq r_{1/2}$, the term multiplying $\kappa_\star$ in 
Equation \ref{eq:approx} is positive and of order unity\footnote{This is because most commonly-used stellar tracer 
distributions yield $\gamma_\star \simeq 3$ at $r_{1/2}$ (see Appendix B of W10) and empirically we expect 
$|\gamma_\sigma| < 1$ owing to the rather flat velocity dispersion profiles observed (see Equations A7-9 of W10).},
which implies that a larger $\kappa_\star$ will favor a smaller $\aleph$ and a more core-like distribution. In the 
limit where $\gamma_\sigma$ and $\gamma_\sigma^\prime$ are small, $\kappa_\star$ will dictate the inferred density slope.

To illustrate this point, consider the light profile of the Carina dSph, which prefers the parameter combination
$\ \rlimit / {\rm r_{_{King}}}$ = 3.27 with a deprojected half-light radius $\rhalf = 334$ pc when fit with a King profile
(see Table 1 of W10). S10 showed that the same photometry can be fit by a stellar light profile that is
divergent at the center (with approximately the same half-light radius).  At $r \simeq \rhalf$ both the S10 fit and the King
fit have similar curvature: $\kappa_\star(\rhalf) = 1.05$ and 1.12, respectively.  But at the slightly smaller radus 
$r =0.5 \, \rhalf$ the light profile favored by S10 has a significantly smaller curvature $\kappa_\star = 1.5$
compared to the King fit $\kappa_\star = 2.4$, and this difference further increases at smaller radii, where 
$\kappa_\star$ diverges at small $r$ for the King profile but drops to zero for the S10 fit. Indeed, compared to 
other common light distributions, the King fit imposes the strongest curvature towards small $r$ and this will act to
push solutions towards more core-like mass density profiles.
For example, a \citet{Plummer11} profile has $\kappa_\star (0.5 \, \rhalf) = 1.4$ and a \citet{Hernquist90} profile has $\kappa_\star (0.5 \, \rhalf) = 0.3$.
Thus, for a fixed $\sigma_r$, King profiles will prefer smaller overall density profile slopes than all of the common alternatives.

The above discussion highlights the difficulties associated with accurately constraining central density profile slopes using Jeans analysis. A similar point was
emphasized by S10. In practice, one would need to know $\kappa_\star$ and $\gamma_\sigma^\prime$ very accurately at small radii in order to definitively constrain the 
asymptotic density profile slope, even with the assumption of isotropy. Our approach instead is to adopt a conservative (and standard) assumption of the King 
tracer distribution, which will allow us to directly compare with the work of G07. We show that even in this limit, not all of the dSph galaxies prefer core-like distributions.

\begin{figure*}
\centering
\includegraphics[width=0.32\textwidth]{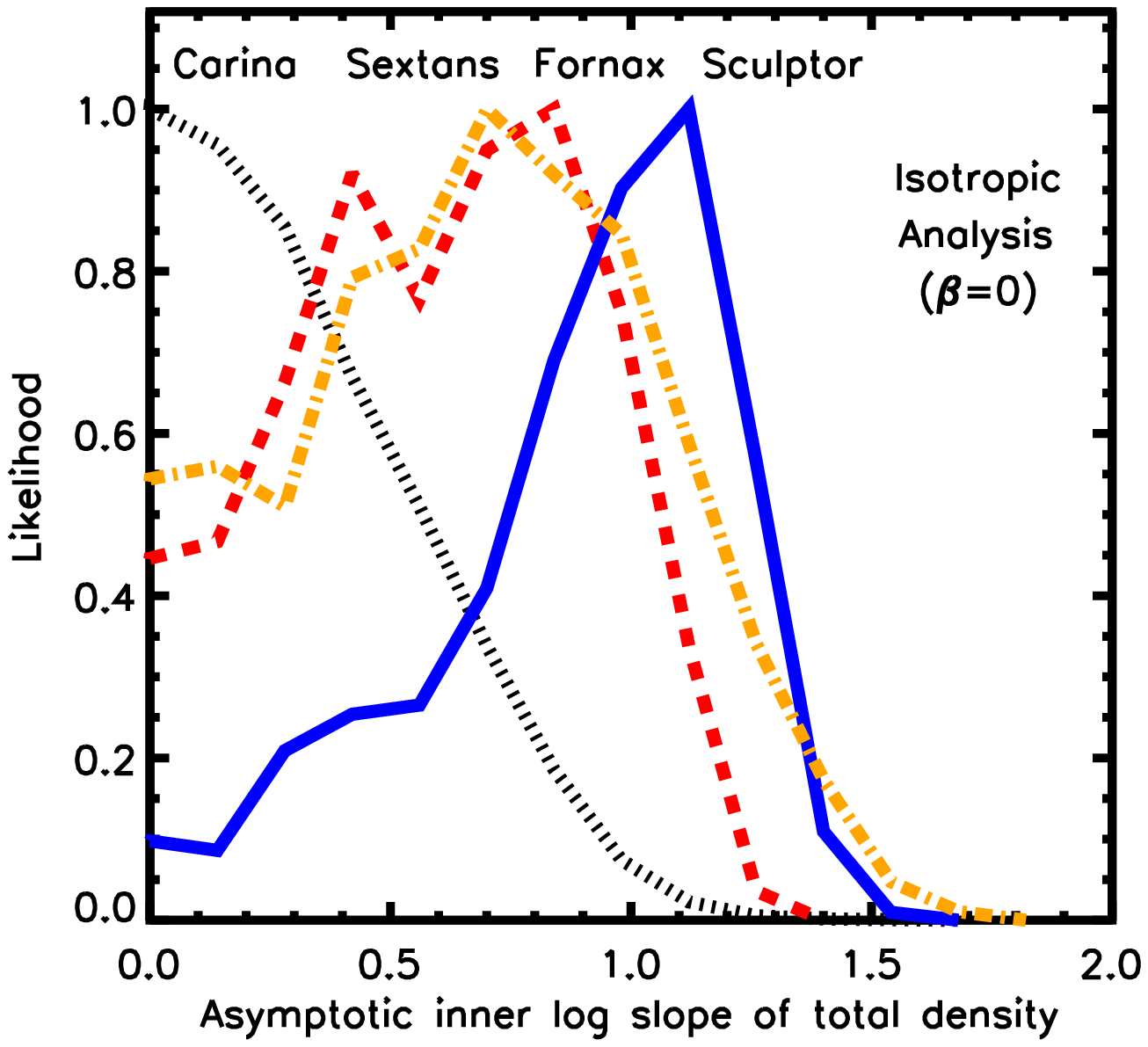}
\includegraphics[width=0.32\textwidth]{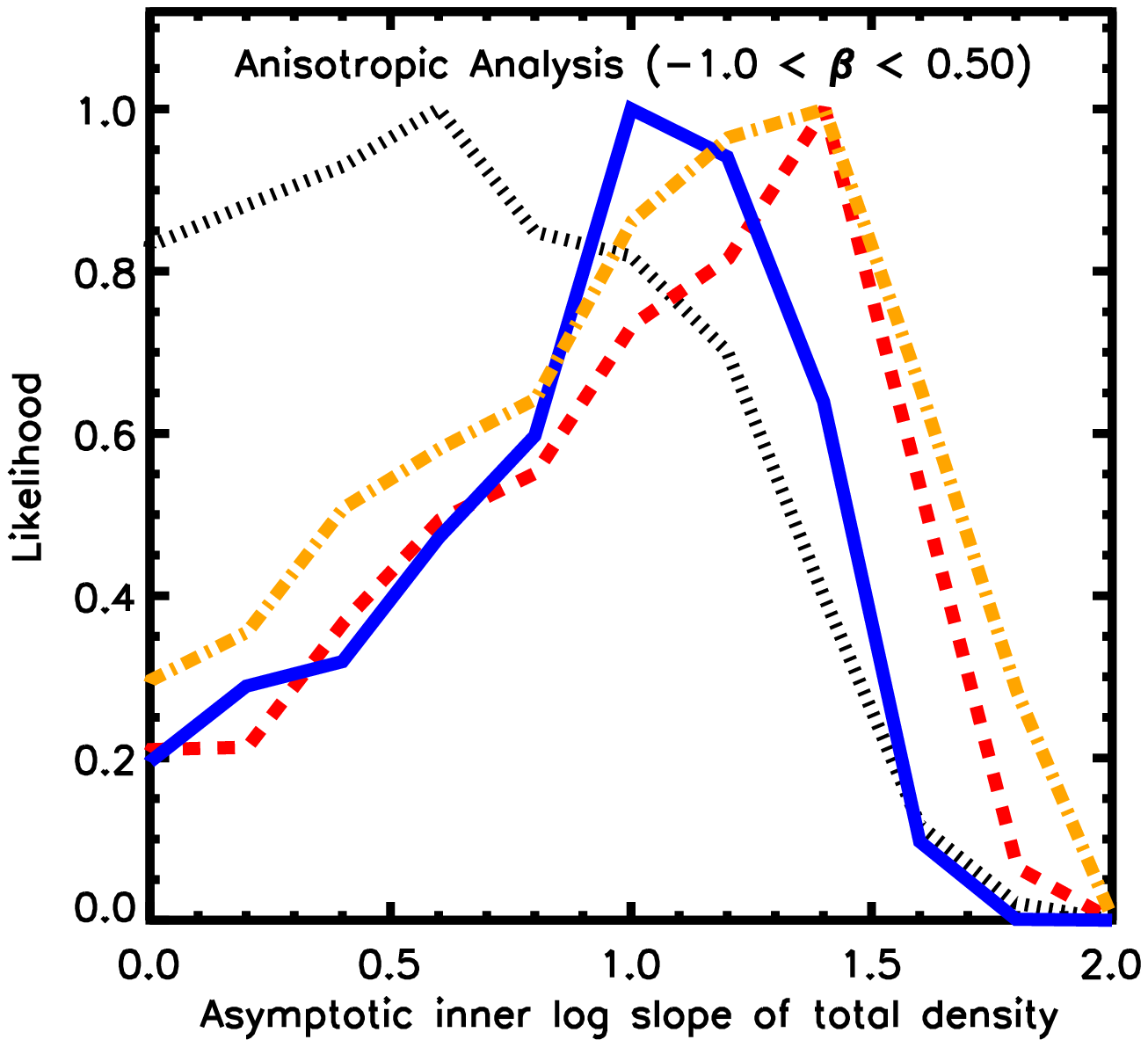}
\includegraphics[width=0.32\textwidth]{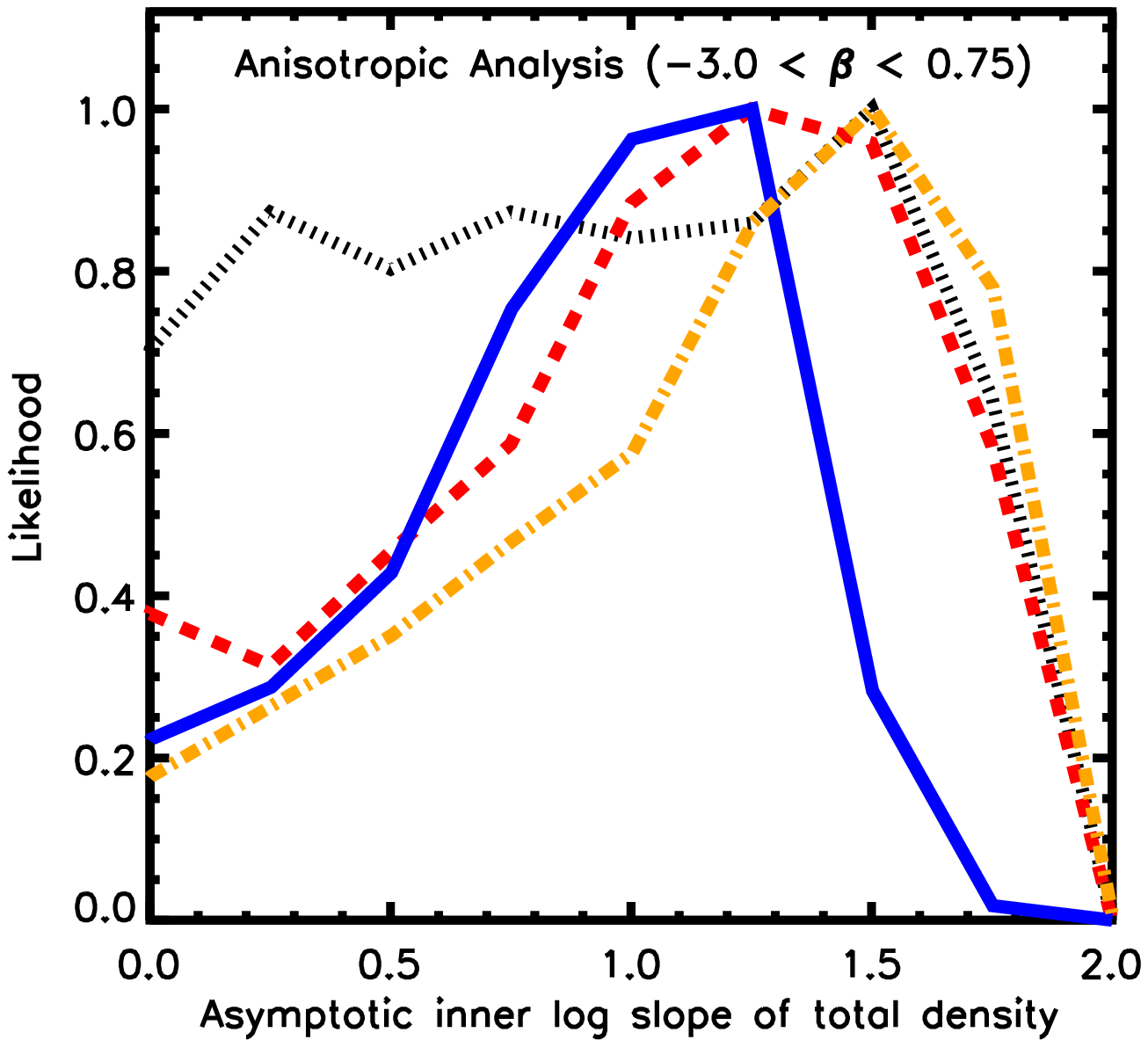}
\caption{Probability densities for $\gamma (r) =  - \rmd \ln \rho / \rmd \ln r$ as $r \rightarrow 0$ when modeling
with Equation \ref{eq:rhor} the four MW dSphs with the highest quality kinematic data which is publicly available:
Carina (black dotted), Fornax (red dashed), Sculptor (blue solid), and Sextans (orange dot-dashed).
{\em (a):} Isotropic analysis ($\beta=0$). This result shows that not
all of the classical MW dSphs prefer to be cored ($\gamma \simeq 0 $) 
under the assumption that $\beta=0$. 
{\em (b) and (c):} Anisotropic analyses demonstrating how the
likelihood for $\gamma$ widens as the range on the allowed
anisotropies widens.} 
\label{fig:likelihood}
\end{figure*} 
%

\section{Dynamical analysis}
\label{subsec:isodyn}
We use a spherical Jeans analyses to constrain the density profile slopes of the eight 
classical MW dSphs: Carina, Draco, Fornax, Leo I, Leo II, Sculptor, Sextans, and 
Ursa Minor. The kinematic and photometric data we use are referenced in Figure 2 and Table 1 of W10
\citep[][]{Munoz05, Koch07, Mateo08, Walker09a}. 
We constrain the allowed mass profile for each dwarf using a Bayesian analysis as 
described in \citet[][]{Martinez09} and W10.  

We marginalize over a generalized density profile described by six parameters
\begin{equation} 
\rho_{\rm tot}(r) = \frac{\rho_s \, e^{-r/\rcut}}{(r/r_s)^c [1+(r/r_s)^a]^{(b-c)/a}} \, ,
\label{eq:rhor}
\end{equation}
using the same prior ranges discussed in Section 2.3 of W10.  When
noted, we have also considered the \citet{Burkert95} profile 
\beq
\rho_{\rm tot}(r) = \frac{\pcore}{(1+r/\rcore)(1+(r/\rcore)^2)}
\label{eq:Burkert}
\eeq
for comparison, with the same above priors for the scale radii and densities. In all of our analyses we marginalize over 
the photometric uncertainties of the fitted \citet{King62} profiles (see Table 1 of W10). Finally, unless otherwise 
stated, we force the velocity dispersion tensor to be isotropic ($\beta = 0$).

\subsection{Density profiles for individual dSphs}
\label{subsec:densityprofiles}
Table 1 summarizes results from our isotropic Jeans analysis for each of the eight galaxies we consider.
The first three columns list relevant observational parameters, while the middle three columns list the derived constraint on the local
density profile slope at three example radii: an asymptotic inner radius ($r \rightarrow 0$ pc), an intermediate radius (100 pc), and
at $\rhalf$. The last two columns display results from our analysis of constraining the density profile by Equation \ref{eq:Burkert}. 
As is shown in the table, and as is highlighted in our discussions below,
some of the galaxies prefer cusps, some prefer cores, and others are indeterminate. Also, as expected from the discussion above, 
galaxies with the smallest curvature in their light profiles $\kappa_\star$ (column 3) tend to be those that prefer the most cuspy overall
density profiles ($\gamma \sim 1$).

Posterior distributions for the asymptotic inner slope 
are presented in Figure \ref{fig:likelihood} for the four dSphs 
that contain the largest published kinematic datasets \citep{Walker09a}.
The left panel shows slope distributions for our isotropic analysis.
Carina (black dotted) demonstrates a strong preference for living in a cored dark
matter halo ($\gamma \simeq 0$), while Fornax (red dashed) and Sextans 
(orange dot-dashed) prefer moderate cusps  ($\gamma \simeq 0.7$), but 
their distributions are too wide to draw any strong conclusions.

The most interesting case is Sculptor (solid blue), which shows a strong preference
for living in a cuspy dark matter 
halo with $\gamma \simeq 1$. Sculptor's line-of-sight velocity dispersion profile rises 
more rapidly in its center than any of the other well-studied MW dSphs, which drives
the data to prefer cuspy profiles when isotropy is assumed. The physical reasoning 
for this is discussed in Section 3.1 of W10. Briefly, as one observes closer
to the center of a system, the intrinsic radial dispersion projects onto the line-of-sight
dispersion. Thus, a sharp rise in the observed dispersion could only be produced by
two effects: the presence of a rising density (i.e. a cusp) or a highly radial stellar
anisotropy. Since we are assuming isotropy in our analysis, a cusp naturally arises.

The middle and right panels of Figure \ref{fig:likelihood} show how the asymptotic 
density slope distributions widen as we allow for wider ranges of (constant) velocity
dispersion anisotropy, $-1.0 < \beta < 0.5$ and  $-3.0 < \beta < 0.75$. Though not
shown here, we also explored a case where the concentrated stellar component of
Fornax was included in the total mass profile with a constant V-band mass-to-light
ratio of $\Upsilon = 2$. Despite the relatively low dynamical mass-to-light ratio of 
this galaxy ($\Upsilon^{\rm V}_{_{1/2}} \simeq 9$; see Table 1 of W10) we did not
find any statistically significant change in the inner dark matter slope distribution.

Figure \ref{fig:likelihood} shows that under the assumption of
isotropy at least one MW dSph (Sculptor) strongly prefers being
hosted in a cuspy dark matter halo.\footnote{Note that G07 
did not analyze Fornax or Sculptor.}
Inspection of Table 1 further reveals that four of the eight classical
dSphs are consistent (one sigma) with inner cusps of $\gamma = 1$,
while three of the  eight dSphs prefer asymptotic slopes of $\gamma
\simeq 0$ at one sigma. 

We also ran an alternate analysis where we set the mass density to be a
Burkert profile (Equation \ref{eq:Burkert}) along the lines of the analysis in 
\citet{Donato09} (hereafter D09)  where it was proposed that over a
wide range of luminosities, galaxies (including the six dSphs in G07)  
can be fit by cored Burkert profiles that have core densities and core radii related 
via $\rho_{\rm core} \propto r_{\rm core}^{-1}$ such that 
$\pcore \, \rcore  = 10^{2.15 \pm 0.2} \, \Msun \, {\rm pc}^{-2}$,
independent of luminosity. Our results (see Table 1) show about a 
factor of four spread in the mean log$_{10}(\pcore \, \rcore)$ values with 
the actual values for log$_{10}(\pcore\, \rcore)$ about a factor of two 
smaller than those in D09. While the product of $\rho_{\rm core}\,\rcore$
is always well constrained, only three dSphs (Fornax, Sculptor, and
Sextans) have constraining posterior distributions for $\rcore$.

Before moving on, we point out that each of these three galaxies have 
$\rcore \simeq \rhalf$. Given that $\gamma(\rcore) = 1.5$ for a Burkert
profile, these results do not indicate disagreement with dissipationless
$\Lambda$CDM simulations, which show that dark matter halos have
similar values for $\gamma$ around the same physical radii.

\begin{figure*}
\centering
\includegraphics[width=0.4985\textwidth]{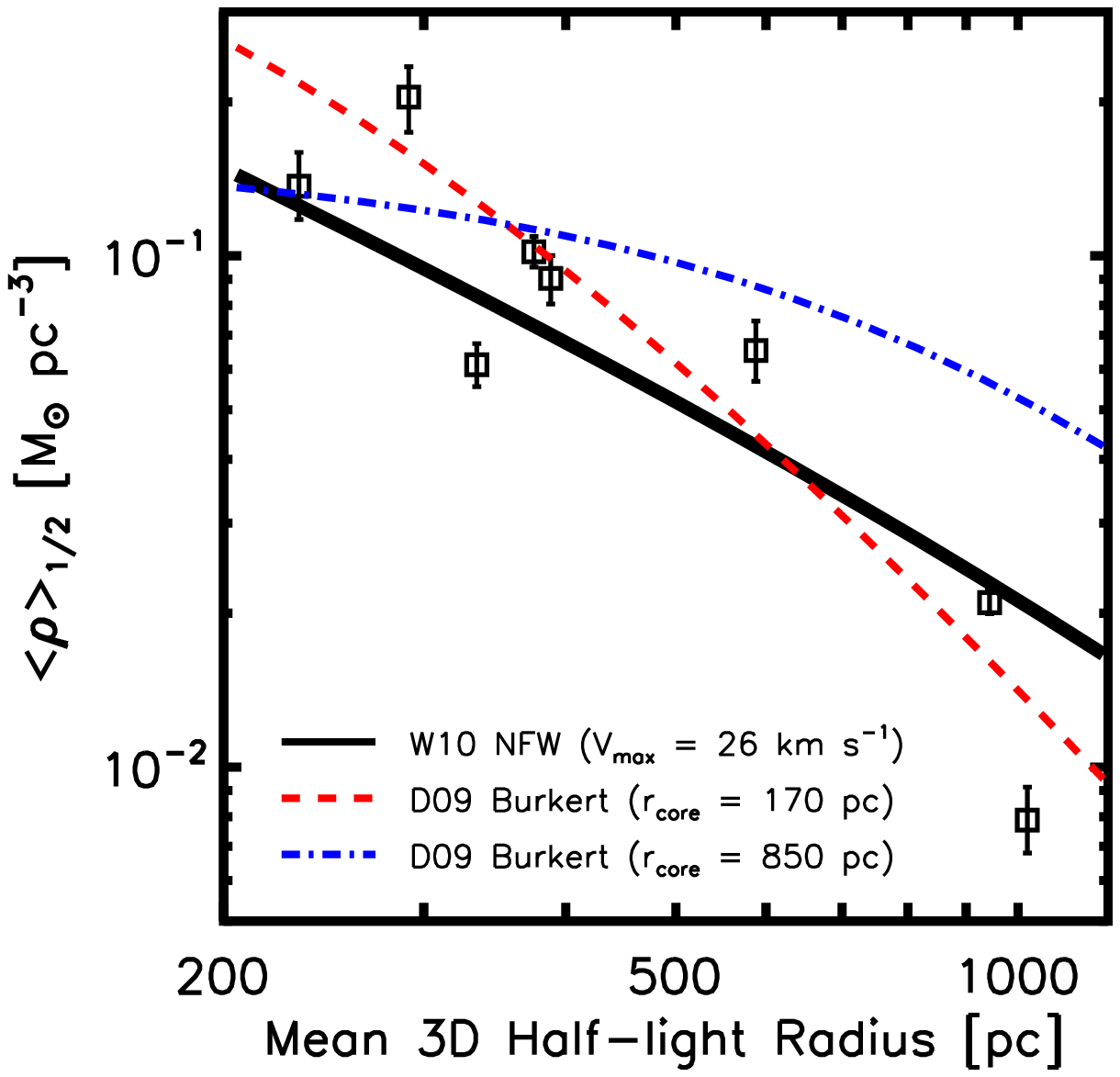}
\includegraphics[width=0.4815\textwidth]{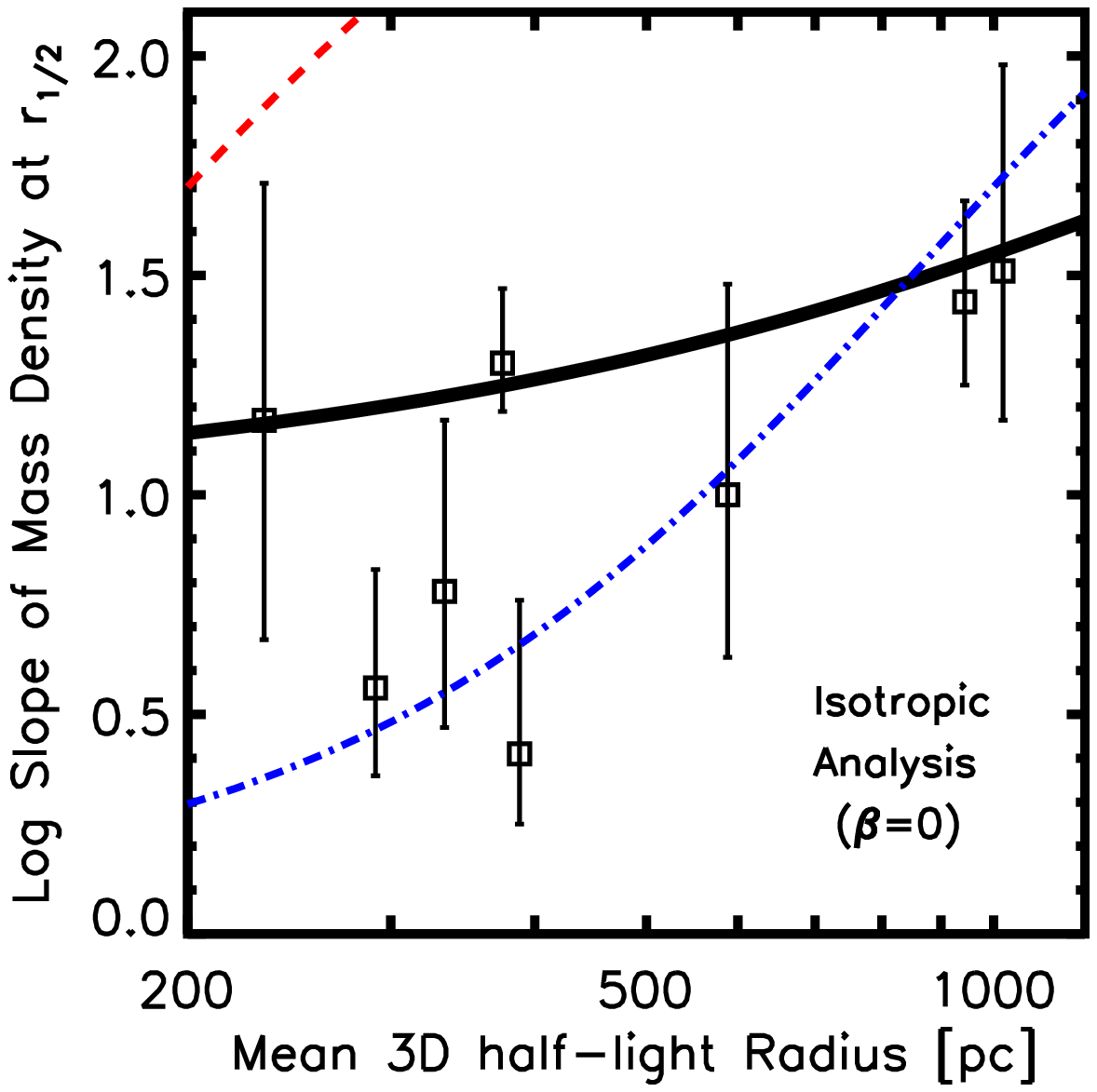}
\caption{Can a common halo reproduce all of the data?  
The solid black lines in each panel show the implied trends for a
typical {\em field} NFW halo with maximum circular velocity 
$\Vmax = 26$ km s$^{-1}$ with $\rmax = 5.7$ kpc (the same NFW halo
presented in Figure 3 of W10). We note that this is less dense than a
typical subhalo of the same $\Vmax$ in recent dark-matter-only
simulations of halos similar to the Milky Way. 
The two colored lines show    
cored Burkert profiles that obey the D09 constant surface density
constraint with core radii of $170$ pc (red dashed) and $850$ pc (blue
dot-dashed). Of these two, only the small cored halo provides a
reasonable characterization of all the data in the left panel, but the
same halo fails to reproduce the log-slope  information in the right
panel for any of the data. Meanwhile, the large cored halo only fares well at
describing some of the data in the right panel. Of the three plotted,  
the cuspy NFW halo fares best at describing the full population within
a common halo. However it is only consistent at the $2-\sigma$ level
with predictions of the $\rmax -\Vmax$ relation from recent
simulations. 
}
\label{fig:rhoslope}
\end{figure*} 

%

\subsection{Global population of MW dSphs}

We now move on to Figure \ref{fig:rhoslope}, where the points presented in the left panel show the
average density of each dSph within its deprojected half-light radius $\langle \rho \rangle_{1/2}$ 
as a function of $\rhalf$. Our motivation for examining the density within $\rhalf$ is that constraints on
the mass density within the {\em deprojected} half-light radius are less sensitive to assumptions about
the velocity dispersion anisotropy, and thus they are more directly constrained by data (W10). The
right panel of Figure \ref{fig:rhoslope} shows the derived local log-slope at $\rhalf$ as a function of
$\rhalf$ for the same set of galaxies under the assumption of isotropy.

Recently, there have been several papers pointing out that the MW dSphs are consistent with inhabiting
similar cuspy dark matter halos \citep[][W10]{Strigari08, Walker09b}.\footnote{Note that \citet{Mateo93} suggested
that the classical dSphs had similar masses within their stellar extent. However, the mass estimator used 
was not accurate (see Appendix C of W10 for an explanation), and thus the claim is not true.}
Additionally, the mass within $\rhalf$ has been shown to increase roughly as $\Mhalf \sim \rhalf^2$
\citep[][W10]{Tollerud10, Walker10}, which is what one would expect
if all of the galaxies were embedded within similar dark matter halos that each obeyed the density scaling $\rho(r) \sim r^{-1}$.  
This is illustrated in the left panel of Figure \ref{fig:rhoslope}, where the solid black line shows a NFW density profile
\citep[which scales as $r^{-1}$ at small $r$; see][]{nfw} for a halo of virial mass $3 \times 10^{9} \Msun$ and a typical
concentration for a {\em field} halo (the same halo shown in Figure 3 of W10, which was
chosen to represent all of the MW dSphs,  not just the classical MW dSphs).

To more accurately assess the hypothesis that all of the MW dSphs lie in 
similar dark matter halos, one needs to simultaneously examine both 
the density {\em and} the log-slope of the population. 
We begin by noting the colored lines in Figure \ref{fig:rhoslope},
which show the implied
$\langle \rho\rangle (<r)$ vs. $r$ and $\gamma (r)$ vs. $r$ relations
for two example cored Burkert profiles -- one with a small core (red
dashed with $\rcore = 170$ pc) and another with a large core (blue
dot-dashed with $\rcore = 850$ pc). The core densities for each of the
Burkert profiles were chosen to follow the previously mentioned
trend advocated by D09.

As can be seen in the left panel of Figure \ref{fig:rhoslope}, in
order to match the density vs. radius relation for the sample of dSphs
we need a Burkert profile that has a core radius that is smaller than
the smallest half-light radius in our sample $r_{\rm core} \lesssim
200$ pc. Although this point was mentioned by \citet{Walker09b}, if we
were to include the ultrafaint MW dSphs as they did, the core size
would need  to be approximately the size of Segue 1 ($\rhalf \simeq
40$ pc; see Table 1 of W10) in order to provide a decent fit to the data.
Moreover, the same small-core Burkert
that best fits the data in the left panel does poorly in the right
panel because it is {\em too steep} to characterize the
population. The large-core Burkert profile (blue dot-dashed) does
better at fitting the data in the right panel but does much worse in
the left panel because the implied core is {\em too
large}. Overall, while it is possible to fit each individual
dSph galaxy with a Burkert profile obeying a scaling of $\rho_{\rm
core} \propto r_{\rm core}^{-1}$ (see Table 1), it is not possible
to find a common cored halo solution that adequately
reproduces the population (i.e. choosing Burkert profiles with 
intermediate core sizes fail in both plots).

However, although it is certainly not a perfect fit to all of the data points in
the right panel, the W10 NFW halo has a local log-slope profile that
is consistent at one sigma with five of the eight dSphs while also 
faring well in the left panel (interestingly, six
of the eight galaxies are one sigma consistent with a log-slope
at $\rhalf$ steeper than $\gamma = 1$). Also note 
that a NFW halo with a scale radius smaller than a kpc 
\citep[like the one advocated for by][]{Walker09b}
will not provide a good fit to the data in the right panel of 
Figure \ref{fig:rhoslope}.

It should be pointed out, however, that the
NFW profile plotted in this figure is lower than the median
predictions from the Aquarius and Via Lactea 2 simulations. 
Thus, Figure \ref{fig:rhoslope} makes it clear that $\beta=0$ is
not a good assumption when attempting to place MW dSphs
in cuspy dark matter halos. We discuss this further
in Section \ref{subsec:LCDM}.

\subsection{Are the solutions physical?}
The solutions obtained by imposing the Jeans equation are not
guaranteed to have a positive definite distribution function. Here we
investigate how a cuspy dark matter density profile is consistent with
$\beta=0$ and $\gamma_\star(r \rightarrow 0)  \rightarrow 0$. Previous
work by \citet{AnEvans06} [also see \citet{White81} and
\citet{deBruijne96} for scale-free potential and stellar profiles] has
shown that the log-slope of the external potential should satisfy an
inequality in order for the distribution function to be positive
definite. The results in \citet{AnEvans06} were noted in the limit
$r\rightarrow 0$, but are applicable everywhere. Here we briefly
recall their arguments.

We assume, as in \citet{AnEvans06}, that the distribution function of
the stars has the form $f(E,L)=L^{-2\beta}g(E)$ and that the dark
matter provides an external potential for this stellar system such
that the mass profile is monotonically decreasing with radius. 
Given the assumption of a constant $\beta$,  
\begin{equation}
n_\star(r) \propto r^{-2\beta} (-\phi(r))^{1/2-\beta} \int_{\phi}^0 (1-E/\phi(r))^{1/2-\beta}
g(E) \rmd E, \label{eq:nstarfromf}
\end{equation}
where $\phi(r)$ is the gravitational potential sourced by the dark
matter density profile, and the constant of proportionality is a
positive definite function of $\beta$. We neglect the influence of
stars on the gravitational potential, which is a good approximation
for the dSphs. With this, it is simple to show that $k(r)=n_\star(r)
r^{2\beta}(-\phi(r))^{\beta-1/2}$ is a monotonically decreasing
function of $\phi$  for $\beta < 1/2$. Since $\rmd \phi(r)/\rmd
r >0$ and $k(r)>0$, we have $\rmd \ln k(r)/\rmd \ln r \leq 0$.    
This constraint implies:
\begin{equation}
\gamma_\star(r) \geq 2 \beta + (1/2-\beta) \delta(r), \label{eq:physical}
\end{equation}
where $-\delta(r) = \rmd \ln (-\phi(r))/ \rmd \ln r >0$ is the log-slope of
the external (i.e. dark matter) gravitational potential. For $\beta=0$,
we have $\delta(r) \leq 2\gamma_\star(r)$. For dark matter density
profiles that diverge more slowly than $1/r^2$ as $r\rightarrow 0$, we
have $\delta(r) \rightarrow 0$. However, this is not
sufficient since we have also assumed that $\gamma_\star(r)
\rightarrow 0$ as $r\rightarrow 0$. 

The log-slope of the potential may be written as: 
\begin{eqnarray}
\delta (r) & = & \frac{V_c^2(r)}{V_c^2(r)+4\pi G \int_r^\infty \rho(x) x \, \rmd x} \\
& \simeq & \frac{V_c^2(r)}{4.5\Vmax^2} \simeq 0.5 (r/r_s)^{2-\gimel}\quad
{\rm as}\; r\rightarrow 0 \nonumber\label{eq:delta}
\end{eqnarray}
where $V_c(r)$ is the circular velocity and $\Vmax$ is the maximum
circular velocity. To derive the final approximation on the second 
line we have assumed that the dark matter density
profile diverges as $(r/r_s)^{-\gimel}$ for $r \ll r_s$. The constants in 
the approximations in the second line are a fit for a NFW
profile but work well for all commonly used cored and cuspy dark
matter halo density profiles. Since $\gamma_\star(r) \propto (r/r_{\rm
King})^2$ as $r \ll r_{\rm King}$, we see that the constraint
$\delta(r) < 2\gamma_\star(r)$ will be violated at small radii  
because $2 \gamma_\star$ for our
tracer density (a King profile) asymptotes to 0 more quickly than
$\delta$ as $r \rightarrow 0$. 

We may estimate the radius at which this constraint is violated by
noting that  $\gamma_\star(r) \simeq 3 (r/r_{\rm King})^2$ for typically
observed values of the concentration of MW dSphs, and hence $\delta(r) >
2\gamma_\star(r)$ for $r\lesssim r_s (0.3 r_{\rm
  King}/r_s)^{2/\gimel}$. For typical values, such as $\gimel=1$, $r_{\rm
  King}=0.3 \, {\rm kpc}$, $r_s=1 \, {\rm kpc}$, the constraint is violated
at radii smaller than about 10 pc. 

Given that both photometric and kinematic data of MW dSphs are
scarce in the region $r \lesssim 10 \, {\rm pc}$ (often the errors on the
position of the  center of the MW dSphs are larger than this), 
substitution of a dark matter distribution with a small core or a 
profile that becomes more shallow with decreasing radius 
\citep[such as the profile presented in][]{Einasto65} would have
little effect on our results outside of this region. In fact, 
recent high-resolution $\LCDM$ simulations do not lend
support to the expectation that the inner density profile log-slope
converges to -1, as an Einasto profile with a gradually decreasing
log-slope amplitude seems to be a better fit 
\citep[e.g.][]{Aquarius, VL2}. Thus, the implications of our isotropic
results would not change were we to fulfill  the requirements of a
positive definite distribution function. 

\begin{figure*}
\centering
\includegraphics[width=0.48\textwidth]{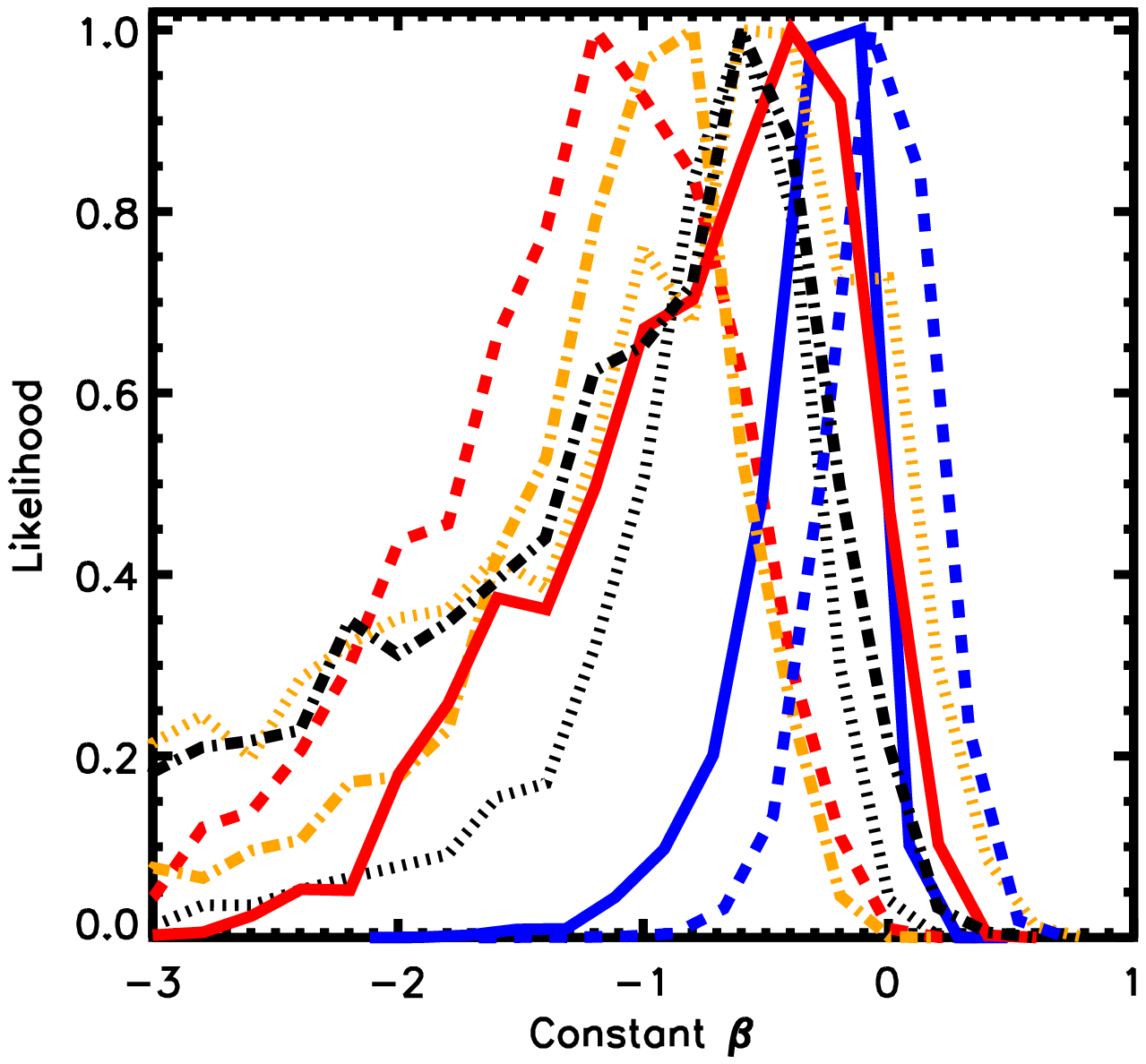}
\hspace{.2in}
\includegraphics[width=0.48\textwidth]{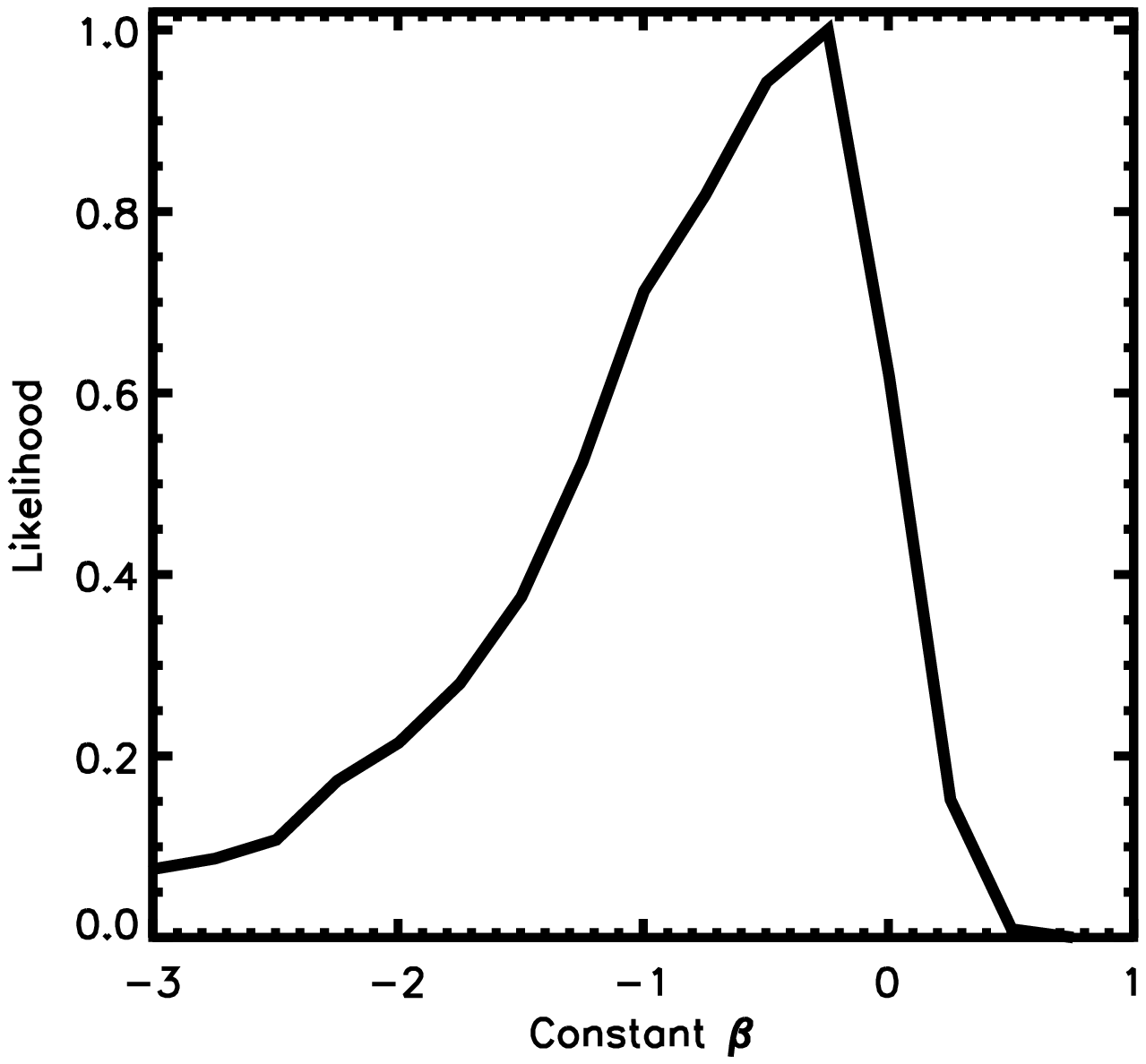}
\caption{
The posterior distribution of anisotropy (assumed to be spatially
constant) for all of the MW dSphs considered in this paper when the density
profile is fixed to be a NFW. The identification scheme is as follows -- 
Carina: black dotted; Draco: red dashed; Fornax: blue solid; 
Leo I: orange dash dotted; Leo II: orange dotted; Sculptor: blue dashed;
Sextans: red solid; and Ursa Minor: black dash dotted.
The plot on the right shows that combined likelihood for $\beta$ 
when each dSph is given equal weight. It is clear that the data demand
a tangential anisotropy in the velocity dispersion tensor for cuspy profile fits.}
\label{fig:betalike}
\end{figure*} 

\subsection{Multiple stellar populations}
Our analysis assumes that the entire spectroscopic
sample follows a spatial distribution that is well described by
the adopted photometric surface brightness fits. We have allowed
for errors in the surface brightness fits, but not 
for the possibility that there are multiple populations with distinct 
velocity dispersions. 

The motivation to model systems with multiple tracer components 
can be understood from the main analytic result of W10,
which shows that the mass within a radius near the
deprojected half-right radius can be determined nearly
independent of stellar velocity dispersion anisotropy. If
two subpopulations have different half-light radii, then one knows
the mass well at two different locations, and thus the effects of the
mass-anisotropy degeneracy are substantially reduced. This
technique was first attempted by \citet[][hereafter  
B08]{Battaglia08} who modeled the Sculptor dSph assuming separate
metal-rich and metal-poor stellar populations. In their analysis, B08
found that Sculptor prefers a cored mass distribution.\footnote{
While this work was being prepared for publication we became aware of
the paper by \citet{AmoriscoEvans12}, which used the B08 data to perform
a similar analysis. Our comments regarding B08 apply also to this work.}

The newer publicly available dataset for Sculptor
\citep{Walker07,Walker09a,Walker09c} that we use contains
approximately three times more velocity measurements than in the B08
sample, and many new spectra are closer to the 
projected center. We emphasize that this larger data set has not been
used to study the issue of multiple populations. 
 Further, B08 impose a strict prior of radial anisotropy (that increases
with distance from the center) in their modeling and this could also
skew the results in favor of cored profiles (see Section
\ref{subsec:LCDM}).  

The most important aspects to address when including information from multiple 
tracer populations in dynamical modeling include both deciding to which 
population each star (for which we have a velocity measurement) belongs
and correctly modeling the underlying radial distributions of the 
subpopulations. For example, in B08 the two
subpopulations are identified in the spectroscopic sample based on a
sharp cut in metallicity (which tends to have a broad distribution in
dSphs) and the radial density profile for each population is assumed to 
follow the radial distribution of the red and
blue horizontal branch stars obtained through
photometry. Uncertainties in this assignment translate to
uncertainties in the estimated velocity dispersion profiles,
especially that of the colder population. In addition, the radial
distribution of the tracer populations should reflect these
uncertainties (and they should not be completely fixed from fits to the
photometric sample). Freedom in the radial distribution of the
subpopulations would imply (as the analytic results of W10 show) that
the locations where the mass is well-constrained become more
uncertain, thus making estimates of the mass profile and its
derivative dependent on assumptions regarding the stellar velocity
anisotropy.

\begin{figure*}
\centering
\includegraphics[width=0.32\textwidth]{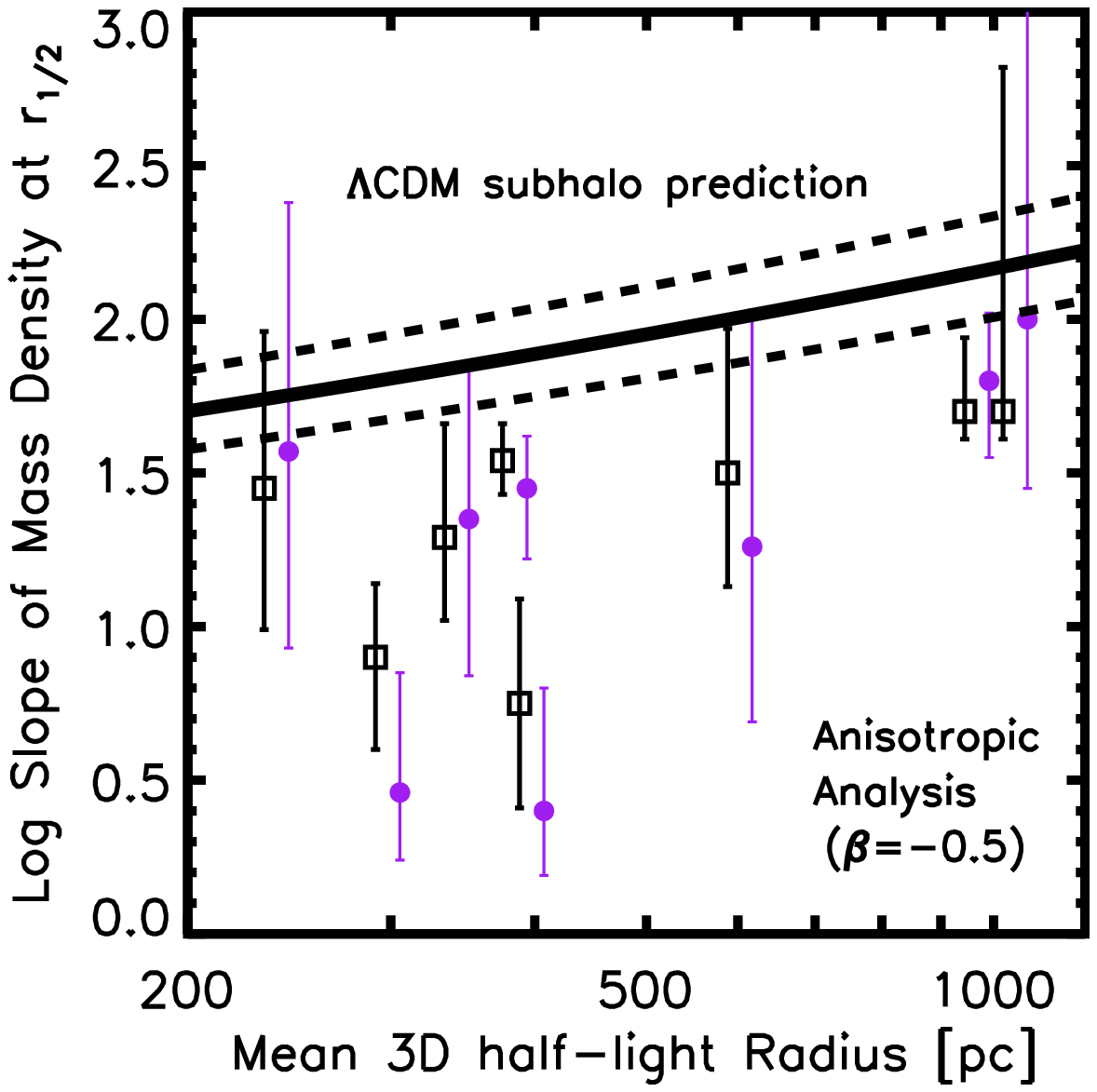}
\includegraphics[width=0.32\textwidth]{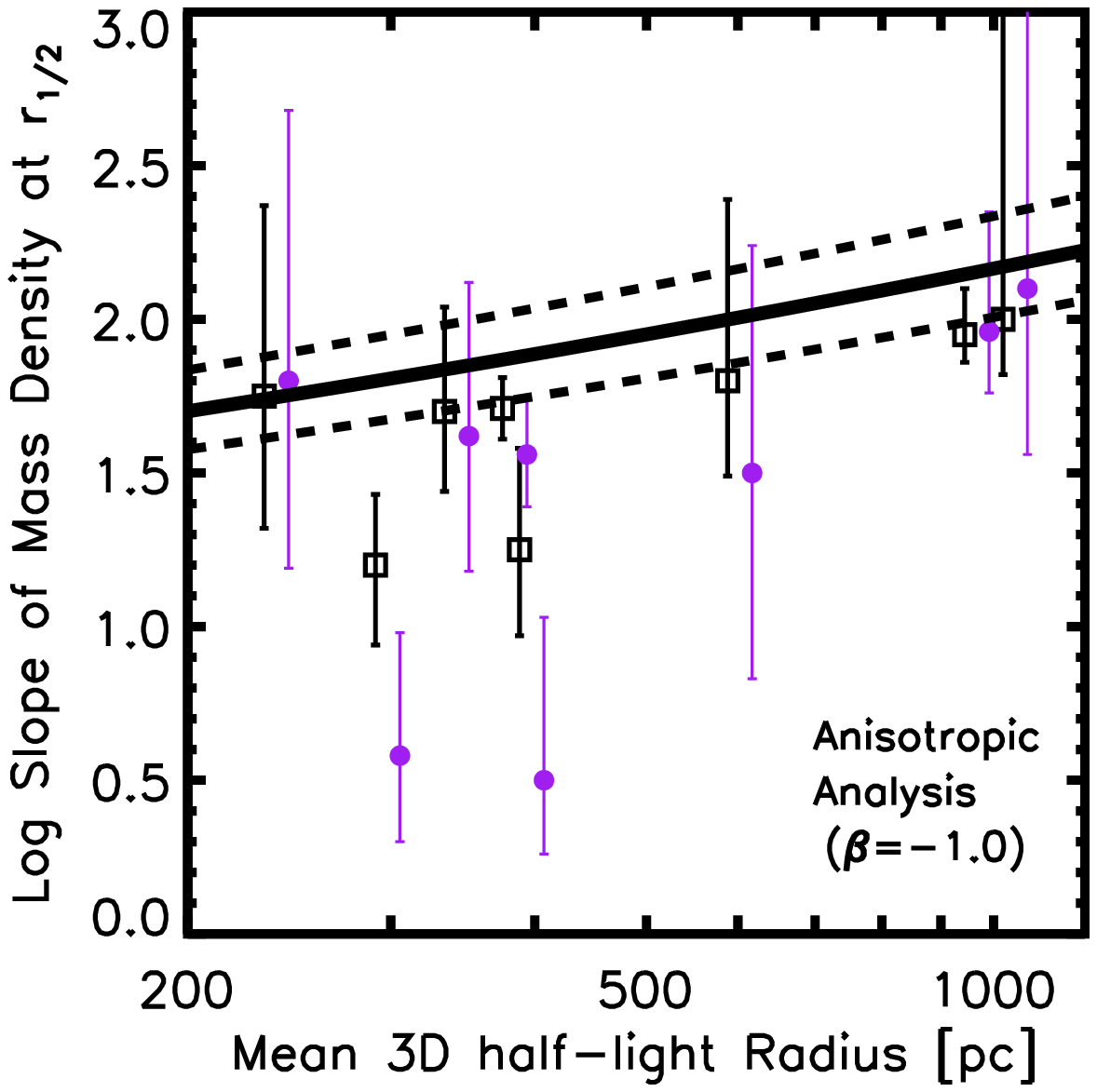}
\includegraphics[width=0.32\textwidth]{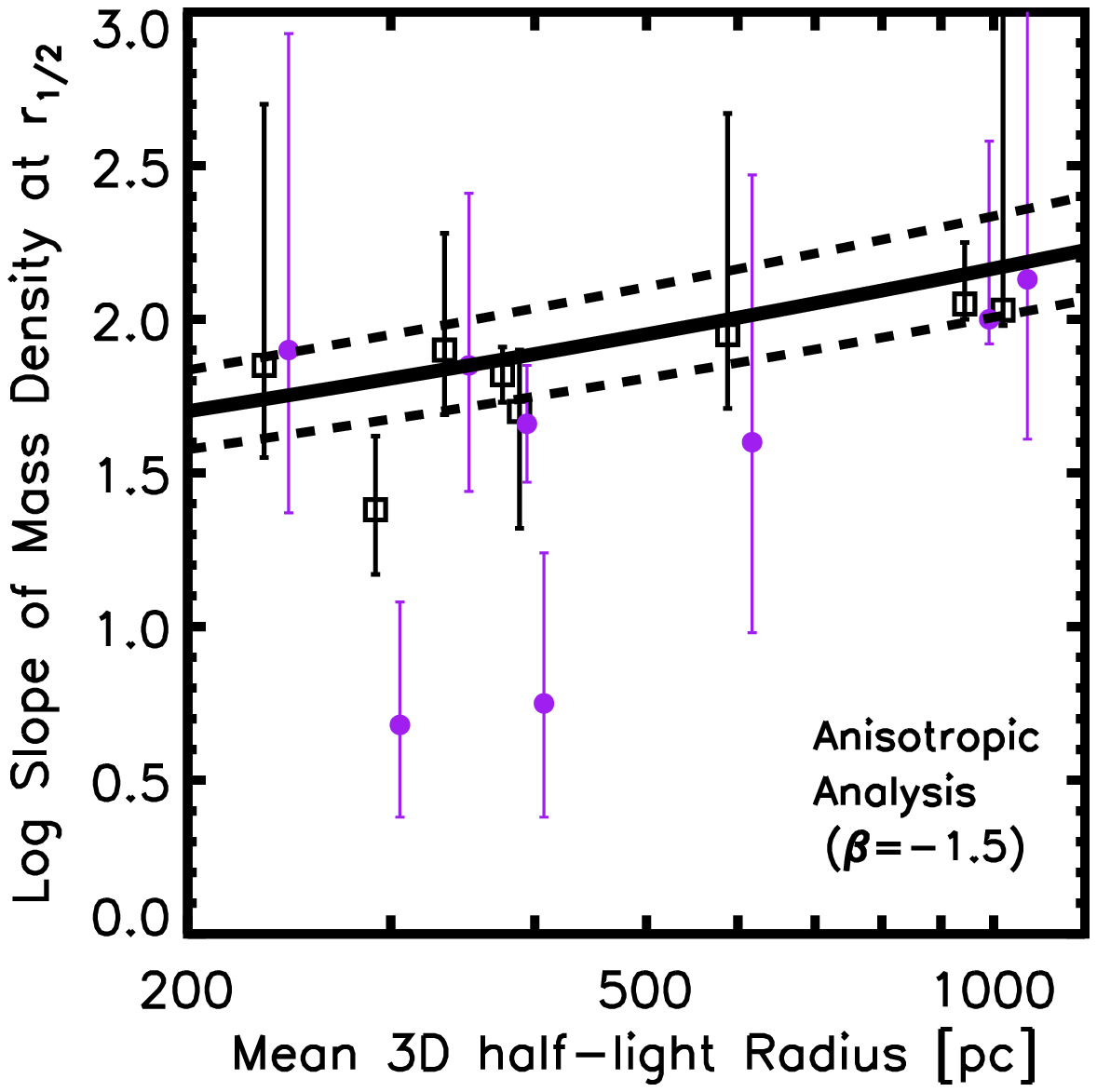}
\caption{
Squares show our derived log-slope of the mass density at the deprojected
half-light radius for each dSph assuming the general density
profile in Equation \ref{eq:rhor}. As marked, each panel presents a different
choice of (tangentially biased) velocity dispersion anisotropy $\beta$. 
The solid black line, which is the same in each panel,
shows the median expectation for $\Vmax=17~{\rm km}/{\rm s}$ subhalos 
in the Aquarius simulation. The dashed lines show the one-sigma deviation
from this relation.
We see that in order for the MW dSphs to be consistent with the
expectations for dissipationless $\LCDM$ subhalos, the stars must
have a substantially larger tangential
velocity dispersion than radial velocity dispersion.
The purple circles, arbitrarily offset to the right for clarity,  show
a test that explores the sensitivity of our results to uncertainties
in the central light profile. For these points, we have removed
data for on-sky radii $R<\rhalf$ and we recomputed the log-slopes.
We see that the overall result is insensitive to the inner data. 
}
\label{fig:nfwfits}
\end{figure*}

A related issue concerns foreground subtraction that may
have benign effects for the entire population as a whole, but could
affect one population more significantly than another.
Further complications arise when one of the populations
is much more centrally concentrated, as is the case for the metal rich
population identified in B08. The isophotal ellipticities could be
different in the inner parts \citep[for Sculptor, see][]{Demers80} and
adopting a half-light radius for each 
population without accounting for these effects could lead to a bias.
A detailed investigation of these issues, following the seminal
work of B08, is required but it is beyond the scope of this paper. 

\subsection{Stellar velocity dispersion anisotropy in a $\LCDM$ context}
\label{subsec:LCDM}

Dissipationless $\LCDM$ simulations predict that dark matter halos generally follow cuspy
density profile distributions over observable scales.
 For concreteness, we
force the dynamical mass to be a NFW profile
($a=1$, $b=3$, $c=1$ in Equation \ref{eq:rhor}) and  then derive the
posterior distributions of $\beta$ implied by the kinematic data.   
Our aim is to gain some insight on values of anisotropy
required of these dSphs should they be embedded in a cuspy halo.

Our results can be seen in Figure \ref{fig:betalike}, where on the left-hand
plot we superimpose the $\beta$ distributions of all eight dSphs. The right-hand
plot shows the combined likelihood of all weighted equally. For most of the dSphs,
tangentially anisotropic velocity dispersions are favored.  The 68\%
range for the combined sample is $-1 \leq \beta \leq 0$.
We note that this range is lower than what is
generally reported for the anisotropy of dark matter particles,
which typically
are close to isotropic in the central regions and 
become radially anisotropic in the outer parts of the halo. However,
there is no theoretical reason to expect the anisotropy of stars to be the
same as that of dark matter.

It is important to
recognize that $\LCDM$ simulations predict not just the radial density
profile of dark matter, but also the range of scale radii expected for
a given subhalo mass or $\Vmax$ (i.e. the subhalo
concentration).  This amounts to a 
prediction for the log slope as a function of physical radius.  
In Figure \ref{fig:betalike} we fixed the density profiles to be NFW but we did not
force the scale radii to match those of subhalos predicted in $\LCDM$ simulations.
Now we aim to investigate whether the derived $\gamma(\rhalf)$ values
are consistent with subhalos of the appropriate concentration for their mass scale
from high-resolution dissipationless $\LCDM$ simulations.

The solid lines in each panel of Figure \ref{fig:nfwfits} show the predicted log-slope
run for a $\Vmax = 17~{\rm km}/{\rm s}$  Einsasto profile halo
with a scale radius set by the median of the Aquarius simulation.
Note that, even though it is not shown, this halo has the appropriate density
normalization to provide a good match to the points in the left panel
of Figure 2. It should be noted that the radial dependence on the predicted log-slope
from $\LCDM$ simulations does not change dramatically for
subhalos that span reasonable $V_{\rm max}$ ranges. 

The black open squares in the three panels of Figure \ref{fig:nfwfits}
show our derived results for $\gamma(\rhalf)$ for three 
choices of anisotropy ($\beta = -0.5, -1.0~\mathrm{and}~-1.5$). In deriving
these points we have allowed a fully general halo profile 
(restoring freedom to all of the variables in Equation \ref{eq:rhor}).
As one moves from left to right in the panels, one can see that the 
only way for the dSphs to be consistent with the median subhalo expectation
is for the velocity dispersion tensor of the stars to 
be  tangentially anisotropic ($\beta \simeq -1.5$).

One concern with this result is that uncertainties in the stellar density profile
shape will hinder our ability to accurately derive $\gamma(\rhalf)$.
The purple circles in Figure 4 (offset arbitrarily to the right for clarity)
show results of a test designed to 
explore this possibility.  Here  we have removed data in the on-sky region
$R < \tfrac{1}{2} \rhalf$.  We found that our results for $\gamma(\rhalf)$ 
did not vary much (i.e. compare the purple circles to the black
squares, which were calculated with all of the data). This implies that the slope of the density
profile at the deprojected half-light radius is determined predominantly
by the local velocity measurements, and thus the uncertainty of the
inner light profile does not contribute significantly to the derived values of
$\gamma(\rhalf)$ for the MW dSphs we are investigating. 

The main point to take from this subsection is that in order
to decently model the classical MW dSphs as galaxies 
living embedded inside of subhalos in $\LCDM$-based
simulations (or even arbitrary NFW halos),
a negative $\beta$ for the stars is usually required.
By imposing $\beta=0$ in the first half of this
paper, we have been stacking the odds against finding cuspy halos. By
simply adjusting all of the anisotropies to mildly negative values
most of the dSphs can live in cuspy halos.  Nevertheless, unless the negative anisotropy
is significant, those
cuspy halos are typically {\em less concentrated} than expected for
subhalos of the appropriate mass. In order to achieve an overall match to 
subhalo scaling relations, fairly significant tangential anisotropies are 
required ($\beta \simeq -1.5$).

\section{Conclusions}
\label{sec:conclusions}

We have performed dynamical Jeans analyses on the eight 
 classical Milky Way dwarf spheroidal galaxies and we have 
 explored their central density structure in comparison
 to predictions for subhalos in dissipationless $\Lambda$CDM simulations.  
We find that in the limiting case of an isotropic velocity dispersion ($\beta = 0$), 
the classical MW dSphs predominantly prefer to live in halos that are less concentrated
 than predicted in these simulations. In 
order to achieve an overall match to the cuspy profile shapes and
subhalo concentrations expected from dissipationless $\LCDM$ simulations,
fairly significant tangential velocity dispersion anisotropy is
required ($\beta \simeq -1.5$).
This result is reminiscent of the well-known central density problem
faced by $\LCDM$ \citep[e.g.][]{Alam02} in comparison to higher mass
galaxies. This effect is also perhaps related to the findings of \cite{BK11, BK12},
who find that the classical MW dSphs are not dense enough to host 
the expected number of massive subhalos around Milky Way size galaxies.
The unexpectedly low concentrations of the dwarf spheroidal galaxies might be 
explained by stochastic feedback effects, which in principle lower the densities 
of dark matter halos compared to what is predicted in dissipationless simulations
\citep[e.g.][]{Mashchenko08, Brook11, Governato10, Pasetto10, Oh11}.
Another possibility is that the dark matter itself is non-standard \citep[e.g.][]{Vogelsberger12}.

Solutions to the isotropic Jeans
equation show that not all of the MW dSphs prefer to live in halos that have
constant density cores at small
radii, even with a conservative treatment of the stellar density
profile. As summarized in Table 1 and Figure \ref{fig:likelihood},
four of the eight dSphs are consistent with $\gamma = 1$ central
cusps at one sigma. Sculptor strongly favors a cuspy inner profile
($\gamma \simeq 1$), while Fornax and Sextans prefer mild cusps
($\gamma \simeq 0.7$), albeit with large uncertainties. Three galaxies
(Carina, Draco, and Leo I) demonstrate a preference for a core
($\gamma \simeq 0$). Therefore, our work lays to rest the possibility
that the assumptions of isotropy and equilibrium force the inner
density profiles of all of the MW dSphs to be cored. Rather,
we have shown that the assumption of isotropy demands that their
{\em concentrations} are lower than
predicted in $\Lambda$CDM, regardless of the presence of a true core in the profile.

Relaxing either the isotropy constraint or the shape of the inner
stellar density profile broadens the allowed range of inner density
slopes. If we adjust all of the anisotropies to negative
values most of the dSphs prefer to live in cuspy halos. 

We also explored whether it was possible to find a common cored dark matter 
halo that simultaneously fits the average local density at $\rhalf$ {\em and} the local log-slope 
at $\rhalf$ under the assumption of isotropy. We showed that such a common halo is not 
possible when restricted to the family of Burkert profiles advocated by D09, but that the
NFW profile used by W10 is a more reasonable fit. If we drop the 
log-slope constraint derived assuming isotropy, then we can find a
single Burkert profile which matches the density-radius relation for
the classical MW dSphs, as long as the core radius is smaller than the
typical size of any galaxy in the population ($\rcore \lesssim 200$
pc). Interestingly, one of the common Burkert halos with which we fit the
local density at $\rhalf$ is actually steeper than the W10 NFW halo
over the radii probed by the data.\\

Acknowledgements $-$ We would like to thank Manoj Kaplinghat, Greg Martinez,
and Louie Strigari for helpful discussions and we acknowledge partial 
support from NASA grant NNX09AD09G.


\end{document}